\documentstyle[aps,epsf,twocolumn]{revtex}

\newcommand{\Ds}{\raise.15ex\hbox{$/$}\kern-.72em\hbox{$D$}}
\newcommand{\be}{\begin{eqnarray}}
\newcommand{\ee}{\end{eqnarray}}
\newcommand{\ben}{\begin{eqnarray*}}
\newcommand{\een}{\end{eqnarray*}}
\newcommand{\ub}{\underline}
\newcommand{\ov}{\overline}
\newcommand{\Gt}{\tilde{G}}
\newcommand{\Db}{\ov{\Delta}_{\pi N}}
\newcommand{\sig}{\Db}
\newcommand{\sigPN}{\sigma_{\pi N}}
\newcommand{\J}[1]{J^{\pi N}_{#1}}
\newcommand{\Jp}[1]{J^{\pi\pi}_{#1}}
\newcommand{\Jn}[1]{J^{NN}_{#1}}
\newcommand{\Gp}[1]{\Gamma^{\pi N}_{#1}}
\newcommand{\Gn}[1]{\Gamma^{N\pi}_{#1}}
\newcommand{\GN}[1]{\Gamma^{NN}_{#1}}
\newcommand{\Gpn}[1]{{\cal G}^{\pi N}_{#1}}

\begin{document}

\draft
\title{\bf On-Shell Approach to Pion-Nucleon Physics} 
\author{{\bf James V. Steele}$^1$, {\bf Hidenaga Yamagishi}
$^2$ and {\bf Ismail Zahed}$^1$}

\address{$^1$Department of Physics, SUNY, Stony Brook, New York 11794,
USA;\\ $^2$4 Chome 11-16-502, Shimomeguro, Meguro, Tokyo, Japan. 153} 
\date{\today}
\maketitle
                                                                               
\begin{abstract}
We discuss an on-shell approach to pion-nucleon physics that is
consistent order by order in a $1/f_{\pi}$ expansion with the
chiral reduction formula, crossing, and relativistic unitarity. 
A number of constraints between the on-shell low-energy parameters 
are derived at tree level in the presence of the pion-nucleon
sigma term, and found to be in fair agreement with experiment. 
We analyze the nucleon form factors, and the $\pi N\to \pi N$ 
scattering amplitude to one-loop, as well as $\pi N\to\pi\pi N$ 
to tree level. We use the latter to derive a new constraint for
the pion-nucleon sigma term at threshold. We compare our results 
to both relativistic and non-relativistic chiral perturbation theory,
and discuss the convergence character of the expansion in light of
experiment.  

\end{abstract}
\narrowtext

\section{Introduction}

Pion-nucleon interactions have been extensively investigated using
dispersion relations and chiral symmetry.  Most of these studies are
built around unphysical points  such as the soft pion limit
\cite{ADLER} or the chiral limit \cite{DASHEN}.
A typical example is the pion-nucleon sigma term --- the
fraction of the nucleon mass due to the explicit breaking of chiral
$SU(2)\times SU(2)$. The scattering amplitude is analytically
continued to the unphysical Cheng-Dashen point \cite{CHENG}, and
chiral perturbation theory (ChPT) is applied \cite{gassnuc}.

An important exception to the above is Weinberg's {\em on-shell}
formula for pion-nucleon scattering \cite{WEINBERG}, which also yields
the Tomozawa-Weinberg relations for the S-wave scattering lengths
\cite{TOMOZAWA}. Recently, we have been able to extend this result to
processes involving an arbitrary number of on-shell pions and nucleons
\cite{MASTER}. A number of identities using the chiral reduction formula
were derived --- one
of which was applied to $\pi\pi$ scattering and shown to be in good
agreement with the data well beyond threshold \cite{pipi}.   

This paper applies the results of the chiral reduction formula
to the nucleon sector, allowing
for an on-shell determination of the pion-nucleon sigma term and $\pi
N$ scattering.  We start by introducing a model in section {\bf II}
that is uniquely specified by the form of the symmetry breaking in QCD
to tree level.  This model can be used to ensure Lorentz invariance,
causality, and positivity while at the same time enforce the constraints
brought about by the chiral reduction formula.
The strategy involved in this calculation compared to
those of ChPT is presented in section {\bf III}.  In section {\bf IV},
we derive an axial Ward identity and discuss the deviation from the
Goldberger-Treiman relation. In section {\bf V}, we recall Weinberg's
relation for $\pi N$ scattering and use the measured S-wave scattering
lengths to predict the pion-nucleon sigma term, the pion-nucleon
coupling, and the induced pseudoscalar coupling to tree level. In
section {\bf VI}, we discuss the one-loop on-shell corrections to the
vector, axial, and scalar form factors, and critically examine the
character of the convergence.  We then evaluate the one-loop
corrections to $\pi N$ scattering in section {\bf VII}.  Finally, we
look at $\pi N \to \pi\pi N$ to tree level and find an additional way
to determine the pion-nucleon sigma term in section {\bf VIII}. Our
conclusions are summarized in section {\bf IX}. Details about the
Feynman rules and the loop expansion are found in the appendices.

\section{Model}

There are two kinds of chiral models possible for the $\pi N$
system. The first is a Skyrme-type model, where the nucleon is a
chiral soliton \cite{SOLITON}. Since solitons often accompany
spontaneous symmetry breaking, this is a natural approach.  Also, if
vector mesons (particularly the omega) are included in chiral
Lagrangians, avoiding soliton solutions is more difficult than having
them.

However, there are two difficulties in this model. One is that the
semiclassical expansion does not commute with the chiral limit. As a
result, the S-wave $\pi N$ scattering lengths are not compatible with
the Tomozawa-Weinberg relation to leading order
\cite{USSOL}. Similarly, the nucleon axial charge $g_A$ is small to
leading order (about half of experiment), and yields a different sign for
$g_A-1$ \cite{USSOL} from that obtained with the Adler-Weisberger sum
rule. This means that a quantitative comparison with experiment is
usually difficult, unless a calculational scheme beyond the
semiclassical expansion is developed.

The other difficulty is more fundamental. In QCD, nucleon operators
$qqq$ and meson operators $\overline{q} q$ exist, which are mutually
local. This has not been shown in Skyrme-type models \cite{USSOL1}.

We will therefore adopt the other type of model, where pions and nucleons are
taken to be independent. 
For the $SU(2)\times SU(2)$ symmetric part of $\pi N$ interactions, we
take the standard non-linear sigma model as the effective Lagrangian
gauged with vector and axial-vector external sources.
\be
{\cal L}_1 &=& \frac{f_{\pi}^2}4 {\rm Tr} \left[ 
 \left(iD_{\mu} U +\left\{ \hat{a}_\mu, U \right\} \right)
\left((iD^{\mu} U)^{\dagger} + \left\{ \hat{a}^\mu, U^\dagger \right\}
\right)   \right]
\nonumber\\
&&+{\overline{\bf\Psi}} \left( i\rlap/\partial + \rlap/{\hat v}
+\rlap/{\hat a}\gamma_5 \right){\bf  \Psi} 
-m_0 \left(\overline{\bf\Psi}_R U {\bf \Psi}_L +
\overline{\bf\Psi}_L U^{\dagger} {\bf \Psi}_R \right)
\nonumber\\
&&+\frac 12 (g_A-1) 
\overline{\bf \Psi}_R 
\left(i\,\Ds U +\left\{ \rlap/{\hat a}, U \right\} \right)U^{\dagger}
{\bf \Psi}_R 
\nonumber\\
&&-\frac 12 (g_A-1) 
\overline{\bf\Psi}_L U^{\dagger} 
\left(i\,\Ds U + \left\{ \rlap/{\hat a}, U \right\} \right) 
{\bf \Psi}_L
\label{1}
\ee
where $U$ is a chiral field, ${\bf \Psi}=({\bf \Psi}_R, {\bf \Psi}_L)$
is the nucleon field, $\rlap/\partial=\gamma^\mu \partial_\mu$, ${\hat
v}_\mu = v_\mu^a \tau^a/2$, and $D_\mu U = \partial_\mu U - i[ {\hat
v}_\mu, U]$.  In the low-energy limit, matrix elements
calculated from (\ref{1}) are essentially unique, given that the
isospin of the nucleon is $\frac 12$ \cite{CCWZ}. Higher derivative (1,1) 
terms at tree level lead to pathologies such as acausality or lack of 
positivity, so they will not be considered.

Ignoring isospin breaking and strong CP violation, the term which
explicitly breaks chiral symmetry must be a scalar-isoscalar. The
simplest non-trivial representation of $SU (2)\times SU (2)$ which
contains such a term is $(2,2)$.  This is the same representation as the
quark mass term $\hat{m}\ov{q}q$ in QCD which generates both the pion
mass and the sigma term.  Therefore we take\footnote{An earlier version of
this work \cite{HEP} used the specific case $c=0$.} 
\be
{\cal L}_2 &=& \frac14 f_\pi^2 m_\pi^2\, {\rm Tr} \left( U + U^\dagger
\right) -\frac{m_\pi^2}{\Lambda}\, \ov{\bf \Psi} {\bf \Psi} 
\nonumber\\
&&{}- c\frac{m_\pi^2}{4\Lambda} \,{\rm Tr} \left( U + U^\dagger
\right) \ov{\bf \Psi}_R U {\bf \Psi}_L + h.c.
\label{2}
\ee
with $c$ and $\Lambda$ arbitrary constants. A bilinear form in ${\bf \Psi}$ 
in (\ref{2}) has been retained, again in analogy with $\hat{m}\overline{q}q$.
We assume that $\Lambda$ is non-vanishing as $m_{\pi}\rightarrow 0$, so that
(\ref{2}) vanishes in the chiral limit.  Scalar and 
pseudoscalar external fields can be added to eq.~(\ref{2}) by taking 
\ben
&&m_\pi^2 {\rm Tr}\, U \to {\rm Tr} \left[(m_{\pi}^2 +s
-i\tau^a p^a)U  \right]  
\\
&&m_\pi^2 \ov{\bf \Psi} {\bf \Psi} \to {\overline{\bf \Psi}} (m_{\pi}^2
+s -i\tau^a p^a \gamma_5 ) {\bf \Psi} 
\een
and similarly for ${\rm Tr}\, U^\dagger$.  The nucleon mass is defined
as $m_N\equiv m_0+\sigma_{\pi N}$ with the pion-nucleon sigma term
$\sigma_{\pi N}=(1+c)m_\pi^2/\Lambda$ as read from the Lagrangian to
tree level. 

Noether's theorem implies that the symmetry breaking term must be
non-derivative, otherwise the vector and axial vector currents will not 
transform as $(3,1)+(1,3)$. The only other term allowable in the $(2,2)$
representation then is $\ov{\bf \Psi}_R U^2 {\bf \Psi}_L + h.c.$ which
is a linear combination of the terms already included in
eq.~(\ref{2}).  Therefore our starting point for the loop-expansion
${\cal L}_{1+2}$ is essentially unique.

The currents used in this paper can be written down by functional
differentiation of the action ${\bf I} \equiv \int\! d^4x\, {\cal
L}_{1+2}$ with respect to the external sources.  In particular, the
pion field is just 
\be
\pi^a (x ) &=& \frac1{f_\pi} \frac{\delta {\bf I}}{\delta p^a(x)}
\nonumber\\ 
&=&-i\frac{f_{\pi}}4 {\rm Tr}\left( \tau^a (U - U^{\dagger}) \right)
+\frac 1{f_{\pi}\Lambda} \overline{\bf \Psi} i\gamma_5\tau^a {\bf \Psi}
\nonumber\\
&&{}+i\frac{c}{4f_\pi\Lambda} {\rm Tr}\left( \tau^a (U - U^{\dagger})
\right) \ov{\bf \Psi}_R U {\bf \Psi}_L + h.c. 
\label{3}
\ee
which reduces to the free incoming pion field $\pi_{\rm in}(x)$ as
$x_0\to-\infty$.  This choice is just the gauge covariant version of
the PCAC pion field, also defined in terms of the axial current ${\bf
A}_{\mu}^a \equiv \delta {\bf I}/\delta a_\mu^a$ as
$\partial^\mu {\bf A}^a_\mu = f_\pi m_\pi^2 \pi^a$.

The one-pion reduced axial current can be defined as the part of the
full axial current that contains no $\pi_{\rm in}$ part. The most 
convenient definition is
\be
{\bf j}_{A\mu}^a &=& {\bf A}_{\mu}^a + f_{\pi} \partial_{\mu} {\pi}^a
\label{4}
\\
&=& g_A\overline{\bf \Psi}\gamma_{\mu} \gamma_5\frac{\tau^a}2 {\bf \Psi} 
+\frac 1{\Lambda} \partial_{\mu}
\left( \overline{\bf \Psi} i\gamma_5 \, \tau^a {\bf \Psi} \right)
\nonumber\\
&&{}-\frac{c}{f_\pi\Lambda} \partial_\mu \left( \pi^a \ov{\bf \Psi}
{\bf \Psi}\right) +{\cal O} (\pi^3) .
\nonumber
\ee
with the expansion to leading order in the PCAC pion field given in
the last two lines. We note that the PCAC pion field is uniquely
defined off-shell within the prescriptions of \cite{MASTER}, and so is
${\bf j}_{A\mu}$. The vector ${\bf V}_\mu^a$  
and scalar $\sigma$ currents are similarly defined through
\ben
{\bf V}^a_{\mu} (x) &=& 
i\frac{f_{\pi}^2}8 {\rm Tr} \left( [\tau^a , U^{\dagger} ] 
\partial_{\mu} U \right) + h.c.
+\overline{\bf \Psi}\gamma_{\mu} \frac{\tau^a}2 {\bf \Psi}
\\
&&{}-\frac 14 (g_A-1)\overline{\bf \Psi}_L \gamma_{\mu} U^{\dagger}
[\tau^a , U ] {\bf \Psi}_L 
\\
&&{}+\frac 14 (g_A-1)\overline{\bf \Psi}_R \gamma_{\mu} 
[\tau^a , U ] U^{\dagger} {\bf \Psi}_R
\een
\ben
\sigma (x ) &=& \frac 1{m_{\pi}^2f_{\pi}} \,{\cal L}_2\, .
\een
The Feynman rules for ${\cal L}_{1+2}$ that are used throughout this
paper are in Appendix A.

\section{Strategy}

We adopt an on-shell loop expansion in $1/f_\pi$ which can be thought
of as a semi-classical expansion with $\sqrt{\hbar}\sim 1/f_\pi$. It
includes pions and nucleon loops beyond tree-level, and is {\it
consistent} order by order with the identities following from the
chiral reduction formula. This
expansion applies equally well to the non-linear sigma-model and QCD
as thoroughly discussed in \cite{MASTER}. We recall that in both
cases, the physical pion decay constant $f_{\pi}$ shows up through the
asymptotic condition of PCAC on the axial-vector current. Therefore,
it is a good expansion parameter when the master formula approach is
applied to these two cases.

All scattering amplitudes will be reduced by the identities 
derived in \cite{MASTER}, and then expanded to one-loop using the
Feynman diagrams from (\ref{1}-\ref{2}). This way, reparameterization
invariance (in the sense of Nishijima-Gursey \cite{GURSEY}) and vector
as well as axial-vector current identities are guaranteed to one-loop. If
we were to just use (\ref{1}-\ref{2}) without the chiral reduction
formula, then $\pi\pi$ scattering subdiagrams in, for example $\pi
N\rightarrow \pi N$ or $\pi N\rightarrow \pi\pi N$ appear to break
reparameterization invariance. In \cite{MASTER} we have checked that
conventional ChPT fulfills the pertinent identities following from
the chiral reduction formula in the mesonic sector. 
We are not aware of such checks in the nucleon
sector\footnote{In \cite{gassnuc} it was shown that Weinberg's
relation for the particular reaction $\pi N\rightarrow \pi N$ holds to
leading order in ChPT, thereby confirming the reparameterization
invariance of their results to the order quoted.}. This work and
others to follow will provide for these checks in our approach.

In our approach broken chiral symmetry and relativistic unitarity will
be addressed for each process individually directly on-shell. This
procedure is conceptually clear, since on-shell renormalization
implies that quantities $m_N$, $g_A$, $\Lambda$, $\sigPN$, $f_{\pi}$,
and $m_{\pi}$ are fixed once and for all at tree level, thereby
including all powers of the quark masses and QCD scale. (In constrast
to ChPT where the chiral logarithms are assessed in these quantities.) 

The ultraviolet finite and non-diagrammatic formulation extensively
discussed in \cite{MASTER} will be presented elsewhere \cite{USBIG}.
To make our exposition in line with current expositions using ChPT, we
will use diagrams. A BPHZ (momentum) subtraction scheme will 
be used throughout. This is to enforce the number of subtraction constants
commensurate with the number of divergences.  Dimensional
regularization is not appropriate, since we need to evaluate nucleon
loops in the axial form factors. These are in general quadratically
divergent, requiring two subtraction constants as opposed to one by
dimensional regularization.

Our strategy is essentially the same as for ordinary renormalizable
theories. No constants other than those required by the divergences will be
considered.  This makes our approach minimal in comparison to ChPT
where {\it all} possible constants required by symmetry and power
counting are used.  This is appealing in that less constants need to
be fixed.  Although ChPT is more general (and generalized ChPT 
\cite{GENERAL} even more so), the excess of constants there 
require additional assumptions such as resonance saturation \cite{TRIO} to 
fix them. In any case, which is the better approach will be dictated by
comparison with experiment.

Below, we will show that to one-loop our results for the form factors
reduce to those obtained by Gasser, Sainio and \v{S}varc
\cite{gassnuc} (GSS) in the context of relativistic chiral
perturbation theory when the nucleon is taken off mass shell
($\Lambda\to\infty$). On mass shell, however, a number of relations
are already observed at tree level in reasonable agreement with
experiment, emphasizing the importance of (broken) chiral
symmetry. What is undoubtedly important in our approach is that the
pion-nucleon sigma term is included at tree level along with the pion
mass term. Since both terms originate from the same quark mass term in
QCD, they naturally go together. This term~--- along with the on-shell
renormalization scheme~--- allow the pions and nucleons to stay on
mass shell to all orders.  A nonzero nucleon scalar form factor and
Goldberger-Treiman discrepancy at tree level are also consequences of
this.

Finally, as is well known, the loop expansion in the pionic sector is
equivalent to a momentum expansion~\cite{gassnuc}. This is no longer
true in the pion-nucleon sector. To overcome this, heavy baryon chiral
perturbation theory (HBChPT) was proposed \cite{MANOHAR}, where an
expansion in $1/m_N$ is made\footnote{Since the nucleon is off mass
shell in these approaches, the expansion is more in terms of a `bare'
nucleon mass.}  \cite{bkm,OTHERS}. However, the relativistic one-loop
calculations contain terms which behave as ${\rm ln}\,m_{\pi}/ m_N$
and are not able to be expanded in this way. Also in this limit
relativistic unitarity is lost, making comparison with experiments
difficult. Lacking a satisfactory theoretical resolution of these
issues, we will maintain a relativistic approach throughout.
Comparison with HBChPT will be made directly by expanding the on-shell
results. 

At this point we note that the unitarity bounds are more stringent
than simple power counting based on a momentum expansion. For instance
in $\pi\pi$ scattering the unitarity bound is saturated for $k^2 \leq
5.2 m_{\pi}^2$ whereas in $\pi N$ scattering the bound is $k^2\leq 3.8
m_{\pi}^2$, indicating in both cases that the expansion parameter
should in fact be closer to $k^2/4\pi f_{\pi}^2$ instead of
$k^2/(4\pi)^2 f_{\pi}^2$.  Throughout, we will work in the kinematical
regime where tree contributions are greater than one-loop, but within
the unitary bounds. All the loop corrections discussed in this work
are on the order of $10-30\%$ of the tree level result with the
exception of the $1/\Lambda$ terms which are small enough to be
sensitive to the input parameters.  These terms will be assessed in as
many ways as possible.

\section{Axial Ward Identity}

The matrix element of the axial-vector current between nucleon states
of momentum $p_i$ and implicit spin dependence $s_i$ can be decomposed
as
\be
&&\langle N(p_2) | {\bf j}^a_{A\mu} (0) |N(p_1) \rangle  =
\nonumber\\
&&=\overline{u} (p_2) \left(\gamma_{\mu} \gamma_5 \,G_1 (t) + 
(p_2-p_1)_{\mu} \gamma_5 \,\overline{G}_2 (t) \right) 
\frac {\tau^a}2 \,\,u(p_1)
\label{5}
\ee
with $t=(p_1-p_2)^2$ and $G_1$ and $\overline{G}_2$ are free of pion
poles. From (\ref{3}-\ref{4}), we also have $\partial^{\mu} {\bf
j}_{A\mu} = f_{\pi} (\Box + m_{\pi}^2 ) \, \pi$. Hence,
\be
&&\langle N(p_2) | \pi^a (0) |N(p_1) \rangle  = 
\nonumber\\
&&= \frac 1{f_{\pi}} \frac 1{m_{\pi}^2-t}
\left( 2m_N\,G_1 (t) + t\,\overline{G}_2 (t) \right)
\overline{u} (p_2)  i\gamma_5\,\frac {\tau^a}2 \,\,u(p_1)
\label{6}
\ee
By definition, eq.~(\ref{6}) is also equal to
\ben
g_{\pi NN} (t) \frac 1{m_{\pi}^2-t} \overline{u} (p_2) i\gamma_5
\tau^a u(p_1)
\een
and leads to the following Ward identity
\be
f_{\pi} g_{\pi NN} ( t )= m_N G_1 (t ) + \frac{t}2
\overline{G}_2 ( t ) 
\label{7}
\ee
where $g_{\pi NN} = g_{\pi NN} (m_{\pi}^2 )$ is the pion-nucleon
coupling constant. Extrapolating from $t=m_{\pi}^2$ to $t=0$ gives the
standard Goldberger-Treiman relation $g_A m_N \sim f_{\pi} g_{\pi
NN}$, with $G_1 (0)\equiv g_A$.

Relation (\ref{7}) is exact.  The Goldberger-Treiman discrepancy is
just given by 
\[
f_\pi\left( g_{\pi NN}(m_\pi^2) - g_{\pi NN}(0) \right) 
\equiv -\Db
\]
We stress again that $g_{\pi NN} (t)$ is physically accessible at both 
$t=m_{\pi}^2$ and $t=0$, making the above discrepancy measurable.
Substituting (\ref{4}) at tree level into (\ref{5}) gives $G_1 (t) =
g_A$ and $\overline{G}_2 (t) = -2/\Lambda$. Therefore using (\ref{7})
we find $\Db=m_\pi^2/\Lambda$ to tree level.  We
choose to renormalize $\Lambda$ on-shell such that this is true to all
orders in the loop expansion.  So
\be
f_\pi g_{\pi NN} = g_Am_N - \frac{m_\pi^2}{\Lambda} \equiv G 
\label{8}
\ee 
and $m_\pi^2/\Lambda$ is exactly the Goldberger-Treiman
discrepancy.

\section{Weinberg's Relation}

One way to determine $G$ and therefore attain a value for $g_{\pi NN}$
is to use pion-nucleon scattering at threshold.  The scattering
amplitude $i{\cal T}$ fulfills a basic Ward identity established by
Weinberg \cite{WEINBERG} and reproduced by the master formula approach
\cite{MASTER}. Taking $(k_1,a)$ as the incoming pion, and $(k_2,b)$ as
the outgoing pion, with $p_1+k_1=p_2+k_2$, the formula is
\ben
i{\cal  T} =  i{\cal T}_V + i{\cal T}_S + i {\cal T}_{AA}
\een
\ben
i{\cal T}_V &=&
-\frac 1{f_{\pi}^2} k_1^{\mu} 
\epsilon^{bac} \langle N(p_2) | {\bf V}_{\mu}^c (0) 
| N(p_1) \rangle_{\rm conn.} 
\\
i{\cal T}_S &=&
-\frac i{f_{\pi}} m_{\pi}^2 \delta^{ab} 
\langle N(p_2 ) | {\sigma} (0) | N(p_1) \rangle _{\rm conn.}
\\
i{\cal T}_{AA} &=&
-\frac1{f_{\pi}^2} k_1^\mu k_2^\nu 
\int\! d^4x\; e^{-ik_1\cdot x}\nonumber\\&&\times
\langle N(p_2) | T^*
{\bf j}_{A\mu}^a (x) {\bf j}_{A\nu}^b (0)| N(p_1)\rangle _{\rm conn.}\,\,.
\een
The isospin structure is decomposed as ${\cal T}^{ba}=\delta^{ab}
{\cal T}^+ + i\epsilon^{bac} \tau^c {\cal T}^-$ to give
\ben
{\cal T}^+ = {\cal T}_S^+ + {\cal T}_{AA}^+
\qquad \qquad
{\cal T}^- = {\cal T}_V^- + {\cal T}_{AA}^-.
\een
The amplitudes ${\cal T}^\pm$ can be calculated to tree level using
the Feynman rules in Appendix A.  At threshold they are
\be
&&{\cal T}^+ = 
\frac{\sigma_{\pi N}}{f_\pi^2} - \frac{\Db^2}{f_\pi^2 m_N} 
-\frac{G^2}{f_\pi^2} \frac{m_\pi^2/m_N}{4m_N^2-m_\pi^2}
\label{9}
\\
&&{\cal T}^- = 
\frac{m_\pi}{2f_\pi^2} \left(1-g_A^2\right)
+ \frac{G^2}{f_\pi^2} \frac{2m_\pi}{4m_N^2-m_\pi^2} .
\label{10}
\ee
Here the pion-nucleon sigma term $\sigma_{\pi N}$ and the
Goldberger-Treiman discrepancy $\Db$ appear.  These
expressions reduce to the Tomozawa-Weinberg formulas for
$\Lambda\to\infty$, showing the corrections are small.

Experimentally, the threshold amplitudes are expressed in terms of the
S-wave scattering lengths ${\cal T}^\pm = 4\pi(1+m_\pi/m_N)a^\pm$.
The Karlsruhe-Helsinki phase shift analysis gives $(a^-,a^+)
=(9.2\pm0.2,-0.8\pm0.4)\times10^{-2}/m_\pi$ \cite{HELSINKI}.  The same
group now has new data from PSI \cite{psi} which reduces many of the
inconsistencies for low pion energies and finds a {\em positive} value
$a^+=0$ --- $4\times10^{-3}/m_\pi$.  Furthermore, pionic atoms give $a^+=
2\pm1\times10^{-3}/m_\pi$ \cite{patom}.  Therefore we take the
weighted mean $a^+=1.5(9)\times10^{-3}/m_\pi$.  The accuracy of
$a^-$ is less of an issue, so we will take the value given above.
These give $\sig=-54\pm10$ MeV and $\sigma_{\pi N}= 14\pm1$ MeV.
Using eq.~(\ref{8}), this value of $G$ gives $g_{\pi
NN}=13.45\pm0.15$, very close to the experimental value $13.4$ taken
in the Paris and Bonn potentials \cite{BON}.  Our on-shell tree level
calculation favors a positive value for $a^+$ and smaller than normal
value for $\sigma_{\pi N}$.  This is opposite to what \cite{bkmpiN}
finds, meriting a one-loop evaluation as carried out in section {\bf
VII}. 

We can also determine the value for the induced pseudoscalar coupling
constant which has been experimentally measured by two groups
\[
g_p \equiv m_\mu G_2(-0.88m_\mu^2) = \left\{ \displaystyle
8.2\pm2.4 \;\;\;\mbox{ref. \cite{CAPTURE}}
\atop\displaystyle 8.7\pm1.9 \;\;\;\mbox{ref. \cite{CAPTURE2}}
\right.
\]
from muon capture in hydrogen.  Using eq.~(\ref{4}-\ref{6}) and the
definition of $G_2$ from the full-axial vector current
\ben
&&\langle N(p_2) | {\bf A}^a_\mu(0) | N(p_1) \rangle =
\\
&&=\overline{u}(p_2) \left( \gamma_\mu \gamma_5 G_1(t) + (p_2-p_1)_\mu
\gamma_5 G_2(t) \right) \frac{\tau^a}2 u(p_1)
\een
gives a relation between $G_2$ and $\overline{G}_2$ 
\be
G_2(t)=\frac{1}{m_\pi^2-t} \left(2m_N G_1(t) + m_\pi^2
\overline{G}_2(t) \right) .
\label{11}
\ee
Using eq.~(\ref{8}) we find to tree level
\[
g_p = \frac{2m_\mu G}{m_\pi^2+0.88m_\mu^2} \simeq 9.0
\]
which is at most $10\%$ higher than the experimental value.

\section{Form Factors}

We now calculate the vector form factor in order to gain insight into
the loop corrections.  It can be decomposed as 
\ben
&&\langle N(p_2) | {\bf V}_\mu^a (0) | N(p_1) \rangle =
\\
&&=\overline{u}(p_2) \left(\gamma_{\mu}F_1(t)+ \frac{i}{2m_N}
\sigma_{\mu\nu}  (p_2-p_1)^\nu F_2(t) \right) \frac{\tau^a}2 u(p_1)
\een
The tree level result just gives the charge
$F_1(t)=1$.  Including the one-loop form of the vector current, we find
\ben
&&F_1(t) = 1
+\frac{2(g_A^2-1)}{f_\pi^2} \left( t c^V_2 + \ov{\ov{\Jp{22}}}(t) \right) 
\\
&&{}+ \frac{G^2}{f_\pi^2} 
\Bigg[2\ov{\Gp{3}}(t) -\ov{\Jn{}}(t) + 8\ov{\Gn{3}}(t) - m_\pi^2
\ov{\Gp{}}(t) 
\nonumber\\
&&{}+8m_N^2 \left( \ov{\Gp{4}}(t) +4\ov{\Gn{4}}(t) 
- 4\ov{\Gn{1}}(t) + \ov{\Gn{}}(t) \right) \Bigg]
\een
\ben
F_2(t)= - \frac{8m_N^2G^2}{f_\pi^2}
\left[ \Gp{4}(t) +\Gn{}(t) +4\Gn{4}(t) - 4\Gn{1}(t) \right]
\een
with the $\Gamma$'s representing loop integrals defined in Appendix A
and $G=g_Am_N-m_\pi^2/\Lambda$ as in the previous section.  An
overlined function denotes a subtraction at $t=0$.  Other than the
subtraction constants, this reduces to the ChPT result of GSS
for $\Lambda\to \infty$.  This is to be expected from the form of the
Lagrangian used and is a good check on the calculation.  Notice that
the coefficients of the terms group into factors of $G/f_\pi\equiv
g_{\pi NN}$ to the order we are calculating by eq.~(\ref{8}).  

Another check is that the charge $F_1(0)=1$ is not renormalized by
loop corrections from the strong force.  This can be shown to be true
in dimensional regularization \cite{gassnuc}.  However, for the rest
of this paper we will instead adopt the BPHZ renormalization scheme.
This just amounts to subtracting the Taylor series of divergent loop
integrals up to the degree of divergence and replacing the subtraction
by arbitrary constants.  This has two advantages: 1) we do not need to
construct the most general Lagrangian to obtain the constants, 2) we
obtain only the minimal amount of constants consistent with the
symmetries of the theory.  In other words, if a diagram is not
divergent, we do not do any subtractions on it.

The only constant to one-loop, $c^V_2$, can be fixed by the vector
charge radius of the nucleon $\langle r^2 \rangle^V_1 =
6F_1'(0)=0.578$ fm$^2$ \cite{hoehler} giving $2(g_A^2-1)c^V_2 =
7.30\times 10^{-3}$.  This accounts for about one third of the radius
with the rest being given by the loop integrals.  This also agrees
with ChPT.

The first difference with ChPT is that there is no subtraction
constant for the Pauli form factor $F_2$.  In fact, the experimental
value for $F_2 (0)$ is just the difference in the anomalous magnetic
moments $\kappa_p-\kappa_n\equiv \kappa_v =3.71$; and a numerical
calculation shows this is indeed valid to about $3\%$ as first shown
by GSS.  The most general Lagrangian, however, allows an extra
subtraction constant for $F_2$  which is not needed \cite{gassnuc}.

We can also check the consistency of our results with those of heavy
baryon chiral perturbation theory by taking $\Lambda\to\infty$ and
expanding in $m_\pi/m_N$ and $t/m_N^2$.  The consistency of the finite
parts of the vector form factor with the relativistic theory has been
shown by \cite{bkm}.  However, the opening of the two pion threshold
at $t=4m_\pi^2$ which is also seen in the data for the vector form
factor \cite{hoehler} is lost in HBChPT.  In addition, extra
divergences in the $1/m_N$ expansion can develop.  If these
divergences cannot be absorbed into the already existing divergence
structure, the procedure of taking the limit of a large nucleon mass
at the level of the Lagrangian could be different than taking this
limit after calculating all amplitudes with a finite nucleon mass.

One such case does occur in the slope of the Pauli form factor
$F_2(t)$.  The full one-loop calculation can be worked out from the
explicit form of the loop integrals in Appendix A to give
($\mu=m_\pi/m_N$)
\ben
&&F_2'(0) = \frac{G^2}{3(4\pi f_\pi)^2} \int_0^1\!\! dy\; \frac{y^3\left[
y^2+4(1-y)^2 \right]}{\left[ y^2+\mu^2(1-y)\right]^2} 
\\
&&= \frac{G^2}{(4\pi f_\pi)^2 m_N^2} \left( \frac{\pi}{3\mu} + 2\ln
\mu^2 + \frac{29}6 \right) + {\cal O}\!(\mu).
\een
which is consistent with GSS in the $\Lambda\to\infty$ limit.  The
$\ln\mu^2$ cannot be expanded in $1/m_N$.  In
principle, this should be taken into account in HBChPT by an
additional subtraction constant.  

Taking $g_A=1.265$ and $\sig=-54$ MeV as in the last section, the
magnetic radius $\langle r^2\rangle_2^V=6F_2'(0)/\kappa_v$ is $0.21$
fm$^2$ to one-loop.  The terms to ${\cal O}(1)$ give $0.31$ fm$^2$ and
the $1/\mu$ term alone gives the HBChPT result of $0.51$ fm$^2$.  The
empirical value is $0.77$ fm$^2$ \cite{hoehler}. Ironically, the first
term in the $\mu$ expansion gives the result closest to experiment.
However, this term can only be singled out in a non-relativistic
expansion which then would require the additional subtraction constant
to absorb the logarithmic singularities mentioned above.

There are no corrections of order $m_N^2/(4\pi f_\pi)^2$ to $F_1(0)$
since it is protected by a non-renormalization condition.  In
addition, $F_2(0)$ had no contribution from tree level and the one-loop
value was shown to be close to the experimental value.  Therefore,
further verdict on the convergence of this expansion requires a
two-loop calculation. 

Extending the results of the previous sections requires the axial-vector
form factors to one-loop.
\ben
&&G_1(t) = g_A -\frac{g_AG^2}{f_\pi^2} \left[ 2 \overline{\Gp{3}}(t) -
\overline{\Jn{}}(t)  - m_\pi^2 \overline{\Gp{}}(t) \right] 
\\
&&\overline{G}_2(t) = -\frac2{\Lambda} + \frac{2g_Am_N G^2}{m_\pi^2
f_\pi^2} \bigg[ 2\ov{\Gp{3}}(m_\pi^2) - \ov{\Jn{}}(m_\pi^2) 
\\
&&\qquad{}- m_\pi^2 \ov{\Gp{}}(m_\pi^2) \bigg]
-\frac{G^2}{f_\pi^2} \Bigg[ 4g_Am_N \ub{\Gp{6}}(t)  
\\
&&\qquad{}+\frac2{\Lambda} \left(\ub{\Jn{}}(t) + m_\pi^2 \ub{\Gp{}}(t)
\right) - 8 I(t) \Bigg] 
\een
with
\ben
I(t) &=& \frac{G}{m_\pi^2-t} \underline{\underline{\Jn{}}}(t) + \frac1{\Lambda}
\underline{\Jn{}}(t)
\een
and an underlined function denotes a subtraction at $t=m_\pi^2$.  The
function $I(t)$ is from the nucleon loop.  It is doubly subtracted in
order to satisfy the consistency relation $\langle0|{\bf
j}_{A\mu}|\pi \rangle=0$ stating that the one-pion reduced
axial current truly does not contain asymptotic pion fields. Both
$G_1(t)$ and $\ov{G}_2(t)$ have subtraction constants which are fixed
by the on-shell renormalization prescription for $g_A$ and $\Lambda$
as discussed at the end of section {\bf IV}. Therefore the one-loop
corrections to $G_1(t)$ are of order $t/(4\pi f_\pi)^2$.  In addition to
this correction, $\ov{G}_2$ has an additional constant correction of
order $\Lambda/m_N$.  We have fixed $\Lambda\sim3m_\pi$ in the last
section.  Therefore this is about a $50\%$ correction to the tree
level result. The corresponding correction to $g_{\pi NN}(t)$ and $G_2(t)$ 
is on the order of a few percent. 

The induced pseudoscalar coupling constant is now
\ben
g_p = \frac{2m_\mu}{m_\pi^2+0.88m_\mu^2} \left[ G
+ \frac{G^2}{f_\pi^2} \left( {\cal F}(-0.88m_\mu^2) - {\cal
F}(m_\pi^2) \right) \right]
\een
with
\ben
{\cal F}(t) &=& G\left[ \ov{\Jn{}}(t) + m_\pi^2 \Gp{}(t) \right] +
4m_\pi^2 I(t)
\\
&&{}-2g_Am_N\left[ \ov{\Gp{3}}(t) + m_\pi^2 \Gp{6}(t) \right]
\een
and $G=g_Am_N-m_\pi^2/\Lambda$ as before.
Taking $m_\mu=106$ MeV, this result gives $g_p\simeq 8.85$ to
one-loop.  This is about a $1\%$ correction to the tree result.  
A plot of $G_2(t)$ for spacelike $t$ is shown in fig.~1.  The dotted
line is the pion pole prediction $2m_NG_1(t)/(m_\pi^2-t)$ with
$G_1(t)=g_A[1-t/M_A^2]^{-2}$ and $M_A=(0.96\pm0.03)$ GeV
\cite{KITIGAKI}. The solid line is from including the one-loop
form for $\overline{G}_2(t)$ for $\sig=-54$ MeV and the
dashed-dotted line is the one-loop form for all of $G_2(t)$.  The
data \cite{DATA} is not precise enough to distinguish between the 
results. 

It should also be noted that taking $\ov{G}_2(t)=0$ in (\ref{11})
along with the linear approximation for $G_1(t)=g_A(1+r_A^2t/6)$
reproduces the Adler-Dothan-Wolfenstein result \cite{adw}
\[
g_p = \frac{2m_{\mu} g_{\pi NN}f_{\pi}}{m_{\pi}^2 + 0.88 m_{\mu}^2}
-\frac 13 g_A m_N m_{\mu} r_A^2
\]
Taking the proper empirical dipole form for $G_1$ gives less than a $1\%$
correction, and including the $\ov{G}_2$ contribution to one-loop
gives about a $5\%$ correction, comparable to the $r_A^2$ term above.
Again more precise data is needed before any statements can be made.

\begin{figure}
\begin{center}
\leavevmode
\epsfxsize=3.3in
\epsffile{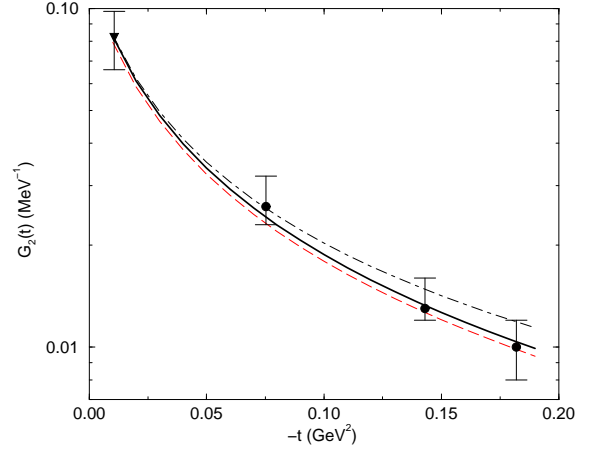}
\end{center}
\caption{The pseudoscalar form factor for spacelike $t$.  The dotted
line is the pion pole prediction and the solid line includes the
one-loop form of $\ov{G}_2$.  The dashed-dotted line is from using the
one-loop form for both $G_1$ and $\ov{G}_2$.}
\end{figure}

Finally, the scalar form factor cannot be directly measured, but is
important in that its value at the Cheng-Dashen point may be tied to
the pion-nucleon scattering data by dispersion analysis
\cite{hoehler}. It is defined as
\[
\langle N(p_2) | \hat{\sigma}(0) | N(p_1) \rangle = 
F_S(t) \overline{u}(p_2) u(p_1) 
\]
\be
F_S(t)&=&-\frac{1+c}{f_\pi\Lambda} -\frac{3}{2f_\pi^3} 
\left(g_A\tilde{G} + \frac{m_\pi^2}{\Lambda} \right) \ov{\Jp{}}(t)
\nonumber\\
&&{}-\frac{3G^2}{f_\pi^3}
\Bigg[ m_N \left( \ov{\Gn{}}(t) - 2\ov{\Gn{1}}(t) \right) 
\nonumber\\
&&\qquad+\frac{1+c}{\Lambda} \left( \ov{\Jn{}}(t) + m_\pi^2
\ov{\Gp{}}(t) \right) \Bigg] 
\label{12}
\ee
The fact that we have kept $\sigma_{\pi N}$ to tree level in the
Lagrangian shows up here as the leading piece in $F_S(t)$.  Since we
renormalize this quantity on-shell, no subtraction constants appear
here. Other than this fact, we agree with GSS for
$\Lambda\to\infty$.  Note that, unlike in the vector form factors, the
coefficients of the terms do not group exclusively into factors of
$G/f_\pi$ showing different factors from just the naive $g_{\pi NN}$.

Defining $\sigma(t)=-m_\pi^2 f_\pi F_S(t)$ as used by other authors
with $\sigma(0)=\sigma_{\pi N}$, we can use eq.~(\ref{12}) to give a
prediction for the scalar form factor at the Cheng-Dashen point 
$t=2m_\pi^2$ \cite{CHENG}. The value of the sigma term obtained
from elastic $\pi N$ scattering (at $t=2m_\pi^2$) as compared to the
value from the baryon mass spectrum (at $t=0$) is about $20$ MeV
larger \cite{SIGMA}.  A numerical evaluation shows for $\sigma_{\pi
N}=20$ MeV that the difference $\sigma(2m_\pi^2)-\sigma(0)=5.3$ MeV
and is not large enough to account for this discrepancy.  This
observation is similar to GSS.

\section{One-Loop $\pi N$ Scattering} 

In order to calculate $\pi N$ scattering to one-loop, we only need
the two-axial-vector correlator to one-loop since the vector and scalar
form factors were evaluated in the previous section.  Defining 
\[
{\cal T}^\pm = A^\pm + \frac12 (\rlap/k_1 + \rlap/k_2) B^\pm ,
\]
and using the Mandelstam variables $s=(p_1+k_1)^2$ and $t=(k_1-k_2)^2$, 
the tree and one-loop result for the form factors can be written as
\ben
A^+_S = \frac{\sigma(t)}{f_\pi^2} &&\qquad
A^-_V = - \frac{s-u}{8f_\pi^2m_N} F_2(t)
\\
&&B^-_V = \frac1{2f_\pi^2} \left[ F_1(t) + F_2(t) \right] .
\een
The rest of the tree result comes from the Born terms of 
$\langle N |T^*{\bf j}_A {\bf j}_A| N \rangle$ .
\ben
A^+_{AA,tree} &=& \frac{g_A\Gt}{f_\pi^2} 
\qquad
A^-_{AA,tree} = 0
\\
B^+_{AA,tree} &=& - \frac{G^2}{f_\pi^2} \left( \frac1{s-m_N^2} -
\frac1{u-m_N^2} \right)
\\
B^-_{AA,tree} &=& - \frac{g_A^2}{2f_\pi^2} - \frac{G^2}{f_\pi^2}
\left( \frac1{s-m_N^2} + \frac1{u-m_N^2} \right)
\een
with $\Gt=2G-g_Am_N$. Its one-loop contribution is quoted in Appendix B.  

Analyzing the divergence structure of the one-loop amplitude shows
that it contains six independent subtraction constants for
the total crossing symmetric amplitude  
\ben
&&f_\pi^4 A^+_{div.} = m_N t\; c_1 + m_N^3\; c_2
\\
f_\pi^4 A^-_{div.} &=& m_N (s-u)\; c_3
\qquad
f_\pi^4 B^+_{div.} = (s-u)\; c_4
\\
&&f_\pi^4 B^-_{div.} = t\; c_5 + m_N^2\; c_6 .
\een
These six constants are in one-to-one correspondence with the six
renormalized constants of GSS: $c^r_{1-4}$, $b^r_2$, and
$b^r_5$.  The five additional finite constants in \cite{gassnuc} have
no counterpart here.  This is a direct consequence of our
minimality assumption: only taking into account the divergent
constants.

The constants may be fixed at subthreshold $s=u$, $t=0$ by fixing the
following constants defined in \cite{hoehler,nagels}
\ben
&&a^+_{00} = (-1.28\pm0.24)/m_\pi 
\qquad 
b^+_{00} = (-3.54\pm0.06)/m_\pi^3
\\
&&a^-_{00} = (-8.83\pm0.10)/m_\pi^2 
\qquad
b^-_{00} = (10.36\pm0.10)/m_\pi^2
\\
&&a^+_{01} = (1.14\pm0.02)/m_\pi^3
\qquad
b^-_{01} = (0.24\pm0.01)/m_\pi^4 .
\een
Since we will be only looking at reactions in the forward direction
($t=0$) below, the coefficients of $t$, $a^+_{01}$ and $b^-_{01}$,
will not be discussed further.  
Using the conventional model for the inclusion of the $\Delta(1232)$
\cite{hoehler} (we take $g_\Delta^2/4\pi=17.7$ GeV$^{-2}$ as in
\cite{kacir} and $Z=\frac12$ as in the original Rarita-Schwinger paper
for the non-pole $\Delta$ terms \cite{RS}), the contribution of the 
$\Delta$ included in the experimental subthreshold values is
\ben
&&a^{+\Delta}_{00} = -1.11/m_\pi  
\qquad
b^{+\Delta}_{00} = -4.86/m_\pi^3 
\\
&&a^{-\Delta}_{00} = -11.34/m_\pi^2 
\qquad
b^{-\Delta}_{00} = 11.62/m_\pi^2 .
\een
One can see that these contributions are large and need to be taken
into account for a proper fit.  We have analytically checked that for any
value of $Z$ and indeed even for the case where the non-pole terms are
{\em neglected} the final result for the scattering lengths presented
below is identical.  The only difference is that part of the strength
of the $\Delta$ contribution is shifted from subthreshold to
threshold; the overall difference between the subthreshold and
threshold remaining the same.  Our choice $Z=1/2$ is merely for
convenience since for this value the $\Delta$ contribution vanishes at
threshold. 

Moving to threshold, a prediction on the scattering lengths can
be made to one-loop.  Taking $g_{\pi NN}=G/f_\pi=13.3$, a reasonable
result is found for $\sigma_{\pi N}=53$ MeV 
\[
a^+_{\rm 1\, loop} \simeq 2.1\times10^{-3}/m_\pi
\qquad
a^-_{\rm 1\, loop} \simeq 12\times10^{-2}/m_\pi .
\]
The value for $a^+$ is close to the weighted average discussed in
section {\bf V} whereas $a^-$ is somewhat
large.  The results are very sensitive to the $\Delta$ contribution.
Taking $\sigma_{\pi N}=45$ MeV gives $a^+ = -8.7\times10^{-3}/m_\pi$
without changing $a^-$.  Taking $\sigma_{\pi N}=0$ and $G=g_Am_N$
give $(a^+,a^-) = (-4.8,10)\times10^{-2}/m_\pi$ similar to
\cite{Gass}.  A more recent calculation \cite{bkmpiN} finds
$a^+=-10\times10^{-3}/m_\pi$.  This shows that the contribution of
the $1/\Lambda$ terms to one-loop really makes a large difference.

Both scattering lengths come from large cancellations between the
constants that were fixed at subthreshold and the loop contribution.
This cancellation is needed due to the close proximity of the tree
result to experiment.  The large contribution from the $\Delta$ clouds
the predictability, but our one-loop analysis seems to favor a value of
$\sigma_{\pi N}$ close to the commonly accepted value of $45\pm8$ MeV
\cite{SIGMA} and a small but positive $a^+$ scattering length.  Both
of these results are in contrast to \cite{bkmpiN} and rely on the
non-zero value of $\sigma_{\pi N}$ and the Goldberger-Treiman
discrepancy at tree level. The value for $a^-$ is about $20\%$ off
from the experimental extrapolation from the Karlsruhe-Helsinki data.

The ability to fix $g_A$ and $g_{\pi NN}$ independently from
experiment allows for a satisfactory starting point to the sensitive
prediction of the $\pi N$ scattering lengths and can also lead to an
estimation of $\sigma_{\pi N}$.  The above analysis, however, showed
an extreme sensitivity of the threshold results in $\pi N$ scattering
and call for further study in the future.

\section{$\pi N \to \pi\pi N$ and the Pion-Nucleon Sigma Term}

As a final estimate on the value of $\sigma_{\pi N}$, we turn to the
process $\pi^a(k_1) N(p_1)\to\pi^b(k_2)\pi^c(k_3) N(p_2)$.  The
scattering amplitude $i{\cal T}_{3\pi}$ fulfills an identity which can
be derived by chiral reduction 
from the master formula approach \cite{MASTER} similar to
what was done for $\pi N$ scattering.  Defining the Mandelstam
variables for a three body process \cite{mandel}    
\ben
&&s=(p_1+k_1)^2 \qquad s_1=(p_2+k_3)^2 \qquad s_2=(k_2+k_3)^2
\\
&&\qquad t_1=(p_1-p_2)^2 \qquad t_2=(k_1-k_2)^2,
\een
the identity is 
\[
i{\cal T}_{3\pi} = \left\{i {\cal T}_\pi + i {\cal
T}_A + i {\cal T}_{SA}  +i {\cal T}_{VA}
\right\} + {\rm 2\ perms}+ i {\cal T}_{AAA}
\]
\ben
&&i{\cal T}_\pi
= \frac{i}{f_\pi^2} (t_2-m_\pi^2)\delta^{ab}  \langle N(p_2) |\pi^c(0)  
|N(p_1) \rangle
\\
&&i{\cal T}_A
= \frac1{2f_\pi^3} (k_2-k_1)^\mu \delta^{ab}  \langle N(p_2) | {\bf
j}^c_{A\mu}(0) | N(p_1) \rangle
\\
&&i{\cal T}_{SA}
=- \frac{im_\pi^2}{f_\pi^2} k_3^\mu \delta^{ab}  \int\! d^4x
e^{-i(k_1-k_2)\cdot x} 
\\
&&\qquad\qquad\times\langle N(p_2) | T^*
\hat{\sigma}(x) {\bf j}_{A\mu}^c(0) | N(p_1) \rangle
\\
&&i{\cal T}_{VA}
= \frac1{f_\pi^3} k_1^\mu k_3^\nu \epsilon^{abe} \int\! d^4x
\; e^{-i(k_1-k_2) \cdot x} 
\\
&&\qquad\qquad\times\langle N(p_2) | T^* {\bf V}_\mu^e(x) {\bf
j}_{A\nu}^c(0) | N(p_1) \rangle 
\\
&&i{\cal T}_{AAA}
=- \frac1{f_\pi^3} k_1^\mu k_2^\nu k_3^\lambda \int\! d^4x_1
d^4x_2\; e^{-ik_1\cdot x_1+ik_2\cdot x_2} 
\\
&&\qquad\qquad\times\langle
N(p_2) |T^* {\bf j}_{A\mu}^a(x_1) {\bf j}_{A\nu}^b(x_2) {\bf
j}_{A\lambda}^c(0) | N(p_1) \rangle
\een
with ``perms'' meaning a permutation of $(k_1,a;$ $-k_2,b;$ $-k_3,c)$.
The structure of the chiral reduction formula immediately shows that the
amplitude depends on $\sigma_{\pi N}$ (through the appearance of the
scalar current $\hat\sigma$ in the ${\cal T}_{SA}$ term) allowing for
an alternative way to fix its value.

At threshold, the amplitude can be decomposed as (see \cite{bkmpipiN}) 
\[
i {\cal T}^{cba} = \frac{\vec{\sigma} \cdot \vec{k_1}}{2m_N} 
\left[ {\cal D}_1 (\tau^b \delta^{ac} + \tau^c \delta^{ab} ) + {\cal
D}_2 \tau^a \delta^{bc} \right] .
\]
The tree contribution to the threshold amplitudes are
\ben
&&f_\pi^3 {\cal D}_1 = \frac32 \left(\frac{m_N+m_\pi}{m_N+2m_\pi}
\right) 
\left[\; \frac{m_N+2m_\pi}{2m_N+m_\pi}\; G + 2\sig \right] 
\\
&&{}- \frac14 (2m_N+m_\pi) \left[ g_A + \frac{3\sig}{m_N+2m_\pi}
\right] 
\\
&&{}+ \sigma_{\pi N} \left[ g_A + \frac{m_\pi
G}{(m_N+m_\pi)(2m_N+m_\pi)} \right] 
\\
&&{}+ \frac14 (2m_N+3m_\pi) \Bigg[ \frac{g_A}4 - \frac{2m_N
G}{(2m_N+m_\pi)(m_N+m_\pi)}
\\
&&\qquad\qquad{}- \frac{3\sig}{m_N+2m_\pi} \Bigg] 
\\
&&{}+\frac{g_A^2}2 (G-2\sig) -\frac{g_A^3m_\pi}2 
\\
&&{}- \frac{G}{(m_N+m_\pi)(2m_N+m_\pi)} 
\Bigg( \frac{g_A^2 m_\pi^2}2 -g_A m_\pi \Gt
\\
&&\qquad\qquad{}+ \frac{2(m_N+2m_\pi)G^2}{2m_N+m_\pi} \Bigg)
\een

\ben
&&f_\pi^3 {\cal D}_2 = -\frac32 \left[\; \frac{m_N+2m_\pi}{2m_N+m_\pi}\;
G+ 2\sig \right]
\\
&&{}+\frac{m_\pi}{2} \left[ g_A + \frac{3\sig}{m_N+2m_\pi} \right]
\\
&&{}+\sigma_{\pi N} \left[ g_A 
- \frac{G}2 \frac{4m_N+5m_\pi}{2m_N^2+2m_Nm_\pi - m_\pi^2} \right]
\\
&&{}-\frac12 (2m_N+3m_\pi) \Bigg[ \frac{g_A}4 - \frac{2m_N
G}{(2m_N+m_\pi)(m_N+m_\pi)}
\\
&&\qquad\qquad{}- \frac{3\sig}{m_N+2m_\pi} \Bigg]
\\
&&{}+\frac{g_A^2}2 (G-2\sig) +g_A^3(m_N+m_\pi)
\\
&&{}- \frac{g_A G \tilde{G}}2 \frac{4m_N+5m_\pi}{2m_N^2 + 2m_N m_\pi -
m_\pi^2} 
\\
&&{}- \frac{2G}{(2m_N+m_\pi)(m_N+m_\pi)} \Bigg( \frac{g_A^2}4 
(6m_N^2+9m_N m_\pi + m_\pi^2)
\\
&&\qquad{}- \frac{G^2}2 \frac{4m_N+5m_\pi}{2m_N+m_\pi} \Bigg) .
\een
The brackets section off, in order, the contribution of the first four
terms in the chiral reduction formula. The contribution from ${\cal T}_{AAA}$
is given by the remaining terms.  This calculation is 
in agreement with \cite{beringer} if we take 
$\Lambda\to\infty$.
Note the explicit dependence on $\sigma_{\pi N}$ in the
threshold amplitudes.  Taking $\sig=-54$ MeV for the proper value of
$g_{\pi NN}$, $\sigma_{\pi N}=53$ MeV as in the previous section, and
defining $D_i={\cal D}_i/2m_N$, we find
\[
D_1 = 2.78 \mbox{\ fm}^3 \qquad D_2 = -6.59 \mbox{\ fm}^3 
\]
whereas experimentally \cite{pipiNexp}
\[
D_1^{\rm exp} = 1.82\pm0.09 \mbox{\ fm}^3 \qquad
D_2^{\rm exp} = -7.30\pm0.24 \mbox{\ fm}^3 .
\]
Although both $D_i$'s depend on $\sigma_{\pi N}$, only $D_1$ is
sensitive to it, decreasing to $D_1=2.5$ fm$^3$ for $\sigma_{\pi
N}=14$ MeV.  Therefore experiment seems to favor a smaller $\sigma_{\pi
N}$.

The large $30\%$ discrepancy in $D_1$ reflects on the difficulty in
extracting the threshold parameters.  A different fit in the
literature \cite{intjourn} gives $(D_1,D_2)=(2.26,-9.05)$
fm$^3$. Furthermore, large corrections occur in ChPT from higher order
terms bringing the convergence of this parameter into question.  The
one-loop corrections to the $D_i$'s will be presented elsewhere.

\section{Conclusions}

We have introduced a minimal model for $\pi N$ dynamics that embodies
uniquely at tree level the main features of broken chiral symmetry
with on-shell pions and nucleons to all orders. Using this on-shell
expansion and a BPHZ subtraction scheme, we have shown how the chiral
reduction formula be enforced with a minimal number of parameters.
All of our results are consistent with data.

With this simple model we have analyzed the axial Ward identity
derived in section {\bf IV} as well as a Ward identity for $\pi N$
scattering originally derived by Weinberg and a new chiral reduction
formula for
$\pi N \to\pi\pi N$.  We have presented a one-loop calculation of the
nucleon form factors and $\pi N$ scattering and shown their
equivalence to ChPT in the $\Lambda\to\infty$ limit.

For finite $\Lambda$, this model has the additional feature of
allowing room for $g_A$, $m_N$, $g_{\pi NN}$, and $\sigma_{\pi N}$ to
be fixed to their phenomenological values --- all at tree level in a
$1/f_\pi$ loop expansion.  In particular, an estimate can be made on
the value of $\sigma_{\pi N}$ from various processes.  Terms
proportional to $\sigma_{\pi N}$ are certainly important in the scalar
form factor $F_S(t)$, the $\pi N$ scattering length $a^+$, and the
threshold parameter $D_1$ from the $\pi N\to\pi\pi N$ process.

The only hindrance in nailing down a more stringent prediction on
$\sigma_{\pi N}$ comes from determining the contribution of nucleonic
resonances such as the $\Delta(1232)$ at the point where the divergent
constants are fixed.  An ideal situation would be to find an amplitude
with the divergences constrained by current conservation and yet still
dependent on $\sigma_{\pi N}$ for an unambiguous determination of the
pion-nucleon sigma term.  Photo-production $\gamma N \to \pi N$ may be
such a case.

On-shell renormalization along with the approach of using a minimal
amount of parameters dictated purely by the divergences increases the
predictability of the model due to fewer constraints needed to
fix the constants.  In particular, there are no constants in $F_2(t)$
or $F_S(t)$ as opposed to one each in ChPT and there are six constants
in $\pi N$ scattering as opposed to eleven in ChPT.  

The analysis of $\pi N$ scattering shows that all six of the
subtraction constants in the amplitude can be fixed at subthreshold.
Including the $\Delta(1232)$ contribution, the scattering lengths can
be predicted with reasonable accuracy.  The value of $\sigma_{\pi N}$
as constrained from $\pi N$ scattering goes from being on the lower
side of the canonically accepted value of $45\pm8$ MeV \cite{SIGMA} at tree
level to within the predicted accuracy at one-loop, showing an improvement
upon adding loop corrections as expected.  

We have kept a relativistic formulation in order to maintain
relativistic unitarity.  Indeed, many HBChPT calculations, although
formulated with a heavy baryon Lagrangian, tend to start with the
relativistic Feynman rules and only after evaluation of the amplitude
take the non-relativistic limit.  This is not only more natural but
keeps from missing terms as could happen from the non-relativistic
formulation.  

The convergence of a relativistic calculation crucially depends on the
appearance of the constant terms $m_N^2/(4\pi f_\pi)^2$.  Although
such terms do appear, in all the cases considered here they are always
accompanied by a divergent subtraction constant.  A mere redefinition
of this arbitrary constant removes such terms from the expansion and
rectifies the convergence.  Whether this is a general feature of the
loop expansion employed here merits further investigation.

\acknowledgements{This work was supported in part by the US DOE grant
DE-FG02-88ER40388.}

\appendix
\newpage

\section{}

The Feynman rules from ${\cal L}_{1+2}$ needed in this paper are ones
with external current lines and internal (loop) pion lines.  We take
the transformation ${\bf \Psi}_i\to \xi_i \psi_i$ ($i=R,L$) with
$U=\xi_R \xi_L^\dagger$ and choose $\xi_R=\xi_L^\dagger$.  The 
rules for the $\psi$ nucleon fields are
\begin{figure}[h]
\begin{center}
\leavevmode
\epsfxsize=3.4in
\epsffile{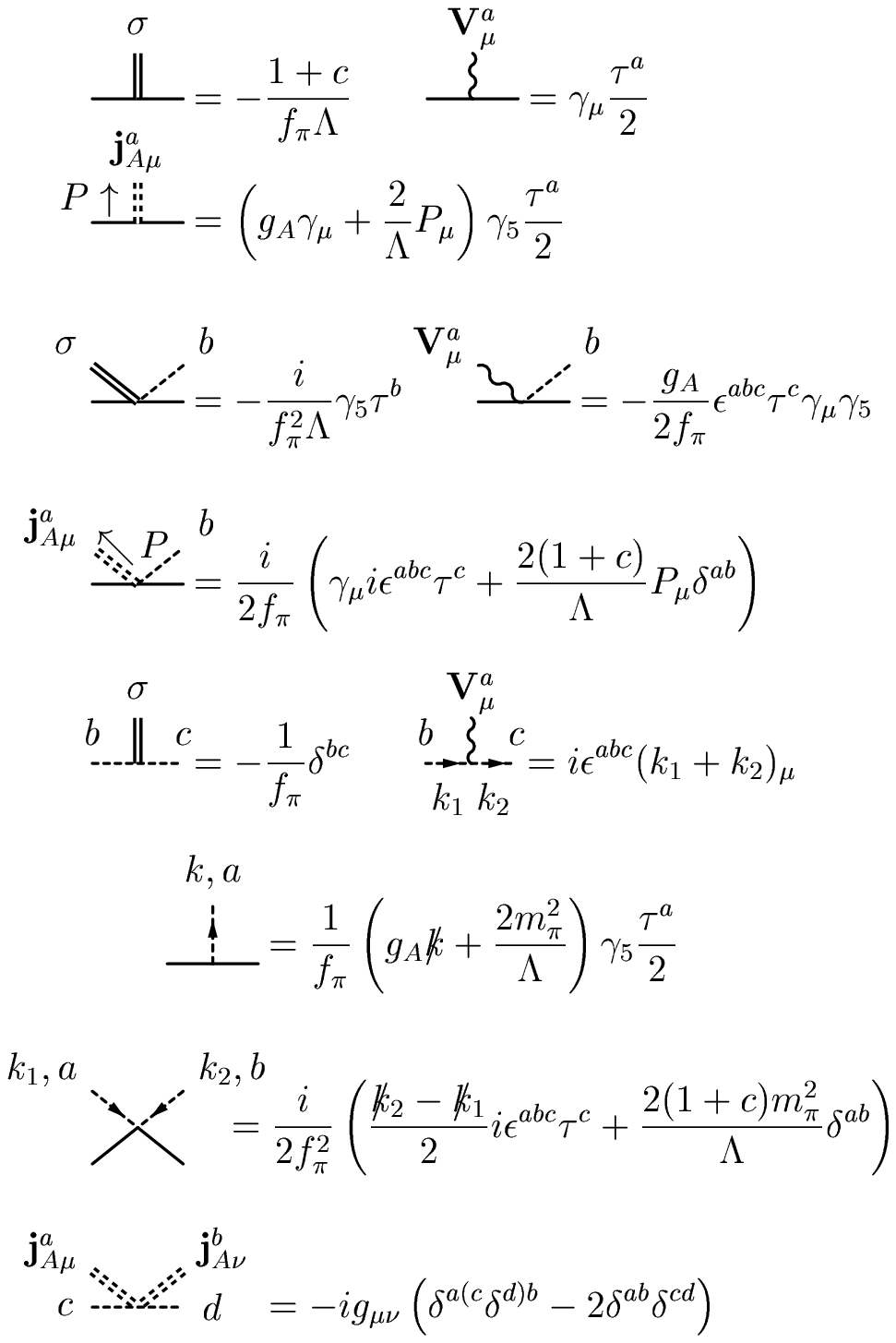}
\end{center}
\end{figure}
All loop processes in this paper can be expressed in terms of the
following general Feynman parameter integrals. 
Here $dk$ is shorthand for $d^4k/(2\pi)^4$ properly
regularized and $[p]_a=(k+p)^2-m_a^2$.  For two propagators, the
integrals are
\ben
-i \int\!\! dk \frac {(1;\; k_\mu;\; k_\mu k_\nu)}{[-p]_a[0]_b}
\equiv \left( J ;\; p_\mu J_1;\; p_\mu p_\nu
J_{21} + g_{\mu\nu} J_{22}\right) 
\een
with $J_i=J^{ab}_i(p^2)$,
\ben
&&\left(\ov{J};\; \ov{J_1};\; \ov{J_{21}} \right) = 
-\int_0^1 \! \frac{dx}{(4\pi)^2} \;\ln \frac{h_J(p^2)}{h_J(p^2_0)} 
\; (1;\; x;\; x^2)
\\
&&\ov{\ov{J_{22}}} = -\!\int_0^1 \!\!\! \frac{dx}{2(4\pi)^2} 
\bigg[ h_J(p^2) \ln \frac{h_J(p^2)}{h_J(p^2_0)} 
+ x(1-x) (p^2-p^2_0) \bigg]
\een
\[
h_J(s) = m_a^2 x + m_b^2 (1-x) - s\, x(1-x),
\]
and $p^2_0$ is the subtraction point. The number of bars above a
function denote how many terms of its Taylor series are subtracted
at the chosen point.  Note that $J^{aa}=2J^{aa}_1$. 

For three propagators, since there are only two particles which play a
role in the loops (nucleon and pion), two of the propagators will
certainly have the same mass.
\ben
&&-i \int \!\! dk \frac{(1;\; k_\mu;\; k_\mu k_\nu; k_\mu k_\nu k_\rho)}
{[0]_a[-p]_b[-q]_b} 
\equiv 
\Bigg( \Gamma ;\; Q_\mu \Gamma_1 + P_\mu \Gamma_2 ;\;
\\
&&\qquad\qquad\qquad
g_{\mu\nu} \Gamma_3 + Q_\mu Q_\nu \Gamma_4 + Q_{(\mu} P_{\nu)}
\Gamma_5 + P_\mu P_\nu \Gamma_6 \Bigg)
\een
with $(Q,P)=(p+q,p-q)$, $\Gamma_i=\Gamma^{ab}(p^2,q^2,P^2)$.  
The finite integrals are 
\ben
\Gamma_i = -\frac1{(4\pi)^2} \int_0^1 \! dx \, dy \; y \; 
\frac{\alpha_i}{h_\Gamma}
\een
with 
\ben
\alpha &=& 1 \qquad 
\alpha_1 = \frac{y}2 \qquad 
\alpha_2 = \frac{y}2 (2x-1)
\\
\alpha_4 &=& \frac{y^2}4 \qquad 
\alpha_5 = \frac{y^2}4 (2x-1) \qquad
\alpha_6 = \frac{y^2}4 (2x-1)^2
\een
and
\ben
h_\Gamma &=& m_a^2 (1-y) + m_b^2 y - p^2 xy(1-y) 
\\
&&{}-q^2(1-x)y(1-y) - P^2 x(1-x) y^2 .
\een
The divergent function can be subtracted to give 
\ben
\ov{\Gamma}_3 = -\frac1{2(4\pi)^2} \int_0^1\! dx \, dy \; y \; 
\ln \frac{h_\Gamma}{h_{\Gamma,0}} 
\een
with $\Gamma_i(m_N^2,m_N^2,P^2) = \Gamma_i(P^2)$ and
$\Gamma_i(s,m_N^2,P^2)=\Gamma_i(s,P^2)$ shorthand notation used in
this paper.  

Finally we need the following functions for four propagators
\ben
&&-i\int\! dk\, \frac{(1;\; k_\mu;\; k_\mu k_\nu)}{[0]_\pi[-p]_N[-q]_N
[-r]_N} \equiv \left( {\cal G};\; p_\mu {\cal G}_1 + q_\mu {\cal G}_2 
+ r_\mu {\cal G}_3 \right)
\een
with ${\cal G}_i = \Gpn{i}(p,q,r)$.  All four integrals are finite
\[
{\cal G}_i  = \frac1{(4\pi)^2}\int_0^1\! dx\, dy\, dz\, y^2z
\frac{\beta_i}{h_{\cal G}^2}
\]
with
\[
\beta = 1\qquad
\beta_1 = xyz \qquad
\beta_2 = (1-x)yz \qquad
\beta_3 = (1-z)y
\]
For this paper only the form $\Gpn{i}(p_1+q_1,p_1,p_2)$ with
$p_1^2=p_2^2=m_N^2$, $p_1+q_1=p_2+q_2$ and $q_1^2=q_2^2=m_\pi^2$ was
used, for which $\Gpn{1}=\Gpn{2}$ and  
\ben
h_G &=& m_\pi^2 (1-y) + m_N^2 y^2 - (s-m_N^2)y(1-y)(1-z)
\\
&&{}- m_\pi^2 y^2z(1-z) -t x(1-x)(yz)^2 
\een
with $s=(p_1+q_1)^2$ and $t=(p_1-p_2)^2$. 

\onecolumn
\section{}

The one-loop form for the two axial-vector correlator 
$\langle N |T^*{\bf j}_A {\bf j}_A| N \rangle$ is quoted
below.  This is needed for $\pi N$ scattering.  Using the notation
$\sig=g_Am_N-G$ we find:

\begin{figure}
\begin{center}
\leavevmode
\epsfxsize=7.2in
\epsffile{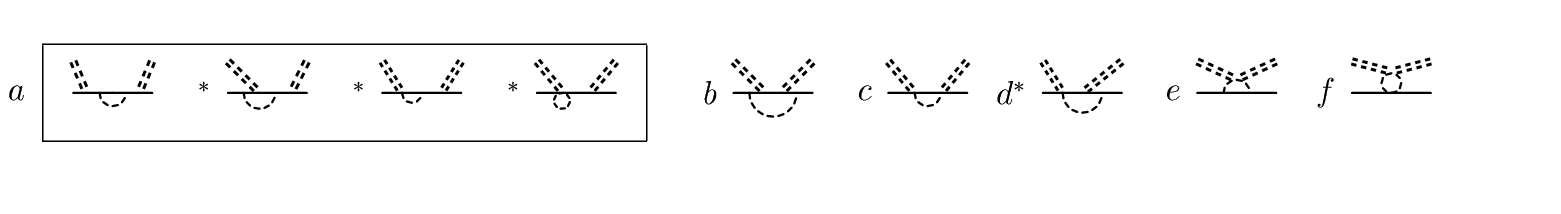}
\end{center}
\caption{\label{piN}
One-loop diagrams for $\pi N$ scattering.  The graphs with a star also
have a mirror image diagram which must be taken into account and all
graphs except for those in the last line require the addition of a
crossed diagram.}
\end{figure}

Writing $A^\pm = A^\pm(s,t,u) \pm A^\pm(u,t,s)$ and $B^\pm =
B^\pm(s,t,u)\mp B^\pm(u,t,s)$ to take into account the crossed
diagrams, we only quote the contribution from the direct diagrams
shown in fig.~\ref{piN}.
The self energy and form factor contributions of fig.~\ref{piN}a can
be written succinctly with the use of the axial-vector nucleon form
factors with one off-shell nucleon leg \cite{USBIG}.  With the
understanding of only taking this to order $1/f_\pi^4$, it can be
written as 
\ben
A^\pm_{\ref{piN}a}(s,t,u) &=& \frac1{4f_\pi^2}
\frac{\sqrt{s}-m_N}{\sqrt{s}} f_{s.e.}(\sqrt{s})
\left[g(\sqrt{s},m_N,m_\pi^2) \right]^2 + (\sqrt{s} \to - \sqrt{s})
\\
B^\pm_{\ref{piN}a}(s,t,u) &=& -\frac1{4f_\pi^2}
\frac1{\sqrt{s}} f_{s.e.}(\sqrt{s}) \left[
g(\sqrt{s},m_N,m_\pi^2)\right]^2  + (\sqrt{s} \to - \sqrt{s})
\een
with $g = (\sqrt{s}+m_N) g_1 + m_\pi^2 \ov{g}_2 - (s-m_N^2) g_3$,
\ben
f_{s.e.}(\sqrt{s}) &=& \frac1{\sqrt{s}-m_N} + \frac{3}{4f_\pi^2}
\frac{(g_A\sqrt{s} + \Gt)^2}{(\sqrt{s}-m_N)^2} \left[ m_N \ov{\J{}}(s)
- \sqrt{s}\, \ov{\J{1}}(s) \right]
-\frac{3m_N^2}{2f_\pi^2} \frac{(g_A\sqrt{s}+\Gt)^2}{\sqrt{s}-m_N}
\left[ \J{}{}'(m_N^2) - \J{1}{}'(m_N^2) \right]
\een
\ben
&&g_1(\sqrt{s}, m_N, t) = g_A 
+\frac{g_A^2-4}{4f_\pi^2} \left(g_A \sqrt{s} + \Gt \right) 
\left[ m_N \ov{\J{}}(s) - \sqrt{s}\, \ov{\J{1}}(s) \right]
\\
&&\qquad{}+\frac{g_AG}{2f_\pi^2} \left( g_A\sqrt{s} + \Gt \right)
\bigg[ -2\ov{\Gp{3}}(s,t) + \ov{\Jn{}}(t) + m_\pi^2 \Gp{}(s,t) 
-(s-m_N^2) \Gp{1}(s,t) - (\sqrt{s} - m_N)^2 \Gp{2}(s,t) \bigg] 
\een
\ben
&&\ov{g}_2(\sqrt{s}, m_N, t)  = -\frac2{\Lambda} 
+\frac{g_A-2(1+c)}{2f_\pi^2\Lambda} \left( g_A \sqrt{s} + \Gt \right)
\left[ m_N \ov{\J{}}(s) - \sqrt{s} \ov{\J{1}}(s) \right]
\\
&&\qquad{}-\frac{G}{f_\pi^2 \Lambda} \left( g_A\sqrt{s} + \Gt \right) 
\left[ \ov{\Jn{}}(t)  + m_\pi^2 \Gp{}(s,t) - (\sqrt{s} - m_N)^2 \Gp{1}(s,t)
-(s -m_N^2) \Gp{2}(s,t)\right]
\\
&&\qquad{}-\frac{g_AG}{2f_\pi^2} \left( g_A\sqrt{s} + \Gt \right)
\left[ (\sqrt{s} - m_N)\left( 2\Gp{5}(s,t) - \Gp{1}(s,t) - \Gp{2}(s,t)
\right) + 2(\sqrt{s} +m_N) \Gp{6}(s,t) \right]
\\
&&\qquad{}+\frac{4G}{f_\pi^2} \left( g_A\sqrt{s} +\Gt \right)
\left[ \frac{G}{m_\pi^2-t} \ub{\ub{\Jn{}}}(t) + \frac1{\Lambda}
\ub{\Jn{}}(t) \right]
\een
\ben
&&g_3(\sqrt{s},m_N,t) = \frac{g_AG}{2f_\pi^2} \left( g_A \sqrt{s} +
\Gt \right) \Bigg[ 2 (\sqrt{s} - m_N) \Gp{4}(s,t) + 2(\sqrt{s} + m_N )
\Gp{5}(s,t)
\\
&&\qquad\qquad\qquad\qquad{}+(\sqrt{s} - m_N) \Gp{1}(s,t) + (\sqrt{s}-
m_N)  \Gp{2}(s,t) \Bigg],
\een
and $\Gt=2G-g_Am_N$. 
It can be checked by the Ward identity in eq.~(\ref{7}) that
$g(m_N,m_N,m_\pi^2) \equiv 2f_\pi g_{\pi NN}$ exactly to one loop
therefore showing a simple way in which the on-shell values are
maintained. The other graphs in fig.~\ref{piN} give
\ben
f_\pi^4 A^+_{\ref{piN}b}(s,t,u) &=& \frac{3}{16} g_A^2m_N \left[
g_A^2(s-m_N^2) + 4\tilde{G}^2 \right] \ov{\J{}}(s)
\\
&&{}+\frac{3}{16} g_A^2 \left[ 2g_As(\tilde{G}-2\sig )
+ m_N\left( g_A^2 s + (\tilde{G}-2\sig)^2 \right) \right]
\ov{\J{1}}(s) 
\\
&&{}+3g_A\Gt G^2 \left[-\ov{\Jn{}}(m_\pi^2) -m_\pi^2 \Gp{}(s) +(s+3m_N^2)
\Gp{1}(s) +(s-m_N^2) \Gp{2}(s) \right]
\\
&&{}+3g_A^2m_NG^2 (s-m_N^2) \Gp{1}(s) +\frac32 g_A^2 G^2 m_N(s-u)
\Gp{4}(t) -6m_NG^4 \Gp{1}(t)  
\\
&&{}+\frac32 g_A\Gt G^2 \left[\ov{\Jn{}}(t) +m_\pi^2 \Gp{}(t) \right] +3m_N
G^4(s-m_N^2) \left[ 2\Gpn{1}+\Gpn{3} \right] 
\\
f_\pi^4 B^+_{\ref{piN}b}(s,t,u) &=&
\frac38 g_A^3m_N (\tilde{G}-2\sig) \ov{\J{}}(s)
+ \frac{3}{16} g_A^2 \left( g_A^2 s + (\tilde{G}-2\sig)^2
\right) \ov{\J{1}}(s) 
\\
&&{}+\frac32 g_A^2 G^2 \left[ -\ov{\Jn{}}(m_\pi^2) - m_\pi^2 \Gp{}(s)
+(s-m_N^2) \left( \Gp{1}(s) + \Gp{2}(s) \right) \right]
\\
&&{}+\frac34 g_A^2 G^2 \left[ 2\ov{\Gp{3}}(t) - \ov{\Jn{}}(t) -
m_\pi^2 \Gp{}(t) \right]
+6g_Am_N\Gt G^2 \Gp{1}(s)
\\
&&{}+3G^4 \left[ (s-m_N^2) \Gpn{3} - m_\pi^2 \Gpn{} -
\GN{}(m_\pi^2,m_\pi^2,t) \right]
\\
A^-_{\ref{piN}b}(s,t,u) &=& -\frac13 A^+_{\ref{piN}b}(s,t,u)
\qquad\qquad\qquad
B^-_{\ref{piN}b}(s,t,u) = -\frac13 B^+_{\ref{piN}b}(s,t,u)
\een
\ben
f_\pi^4 A^+_{\ref{piN}c}(s,t,u) &=& \frac12 m_N
\left[ (s-m_N^2+2\sigma_{\pi N}^2) \ov{\J{}}(s) -(s-m_N^2-2\sigma_{\pi
N}^2) \ov{\J{1}}(s) \right] 
\\
f_\pi^4 A^-_{\ref{piN}c}(s,t,u) &=& \frac14 (s-m_N^2) 
\left[ m_N \ov{\J{}}(s) - (m_N-4\sigma_{\pi N}) \ov{\J{1}}(s) \right]
\\
f_\pi^4 B^+_{\ref{piN}c}(s,t,u) &=& -\frac12 \left[ 2m_N^2
\ov{\J{}}(s)  -(s+m_N^2+2\sigma_{\pi N}^2) \ov{\J{1}}(s) \right] 
\\
f_\pi^4 B^-_{\ref{piN}c}(s,t,u) &=& -\frac14 \left[
2m_N(m_N-2\sigma_{\pi N}) 
\ov{\J{}}(s) - (s+m_N^2-4m_N\sigma_{\pi N}) \ov{\J{1}}(s) \right] 
\een

\ben
f_\pi^4 A^+_{\ref{piN}d}(s,t,u) &=& -\frac12 g_Am_N \left(
g_A(s-m_N^2) -2\sigma_{\pi N} \Gt \right) \ov{\J{}}(s)
\\
&&{}-\frac12 g_A\left( (g_Am_N-4\sig) (s-m_N^2) - g_A \sigma_{\pi N} (s+m_N^2)
+4m_N\sig \sigma_{\pi N} \right)  \ov{\J{1}}(s)
\\
&&{}+2\sigma_{\pi N} G^2 (s-m_N^2) \Gp{2}(s) -2\sigma_{\pi N} G^2 
\left(\ov{\Jn{}}(m_\pi^2) + m_\pi^2 \Gp{}(s) \right) 
\\
f_\pi^4 B^+_{\ref{piN}d}(s,t,u) &=& \frac12 g_Am_N \left(
g_A\sigma_{\pi N} + 4\sig\right) \ov{\J{}}(s) 
\\
&&{}-\frac12 g_A\left( g_A ( s-m_N^2 -m_N \sigma_{\pi N}) +
2\sig (2m_N+\sigma_{\pi N}) \right) \ov{\J{1}}(s)
\\
&&{}+ 2G^2 \left( \ov{\Jn{}}(m_\pi^2) + m_\pi^2 \Gp{}(s) \right) 
\\
&&{}-2G^2 (s-m_N^2) \left( \Gp{1}(s) + \Gp{2}(s) \right) +4m_N
\sigma_{\pi N} G^2 \Gp{1}(s)
\\
A^-_{\ref{piN}d}(s,t,u) &=& 0
\qquad\qquad\qquad
B^-_{\ref{piN}d}(s,t,u) = 0
\een
\ben
f_\pi^4 A^+_{\ref{piN}e}(s,t,u) &=& (t-2m_\pi^2) \left[
g_A\Gt \ov{\Jp{}}(t) + 2m_N G^2 \left( \Gp{}(t) - 2\Gp{1}(t) \right) \right]
\een
\ben
f_\pi^4 A^+_{\ref{piN}f}(s,t,u) &=& \sigma_{\pi N} (t-2m_\pi^2) \ov{\Jp{}}(t)
\een
In the $\Lambda\to\infty$ limit we reproduce all the finite parts of
GSS. This shows that calculation of the $\pi N$ scattering amplitude
using the Ward identity with external fields is equivalent to a
calculation with the pion vertices from the Lagrangian without mention
of external fields or the Ward identity. Whether this holds for $\pi N
\to\pi\pi N$ will be discussed elsewhere.

The relation between the two calculations can be made more transparent
by use of diagrams.  The vector and scalar form factor
contribution to the 
$\pi N$ Ward identity, along with the contact interactions from
$\langle N | T^* {\bf j}_A {\bf j}_A | N \rangle$, are equivalent to the
the graphs from the pion calculation which contain two external pions
meeting at one point. Diagramatically this is just:

\begin{figure}
\begin{center}
\leavevmode
\epsfxsize=7.2in
\epsffile{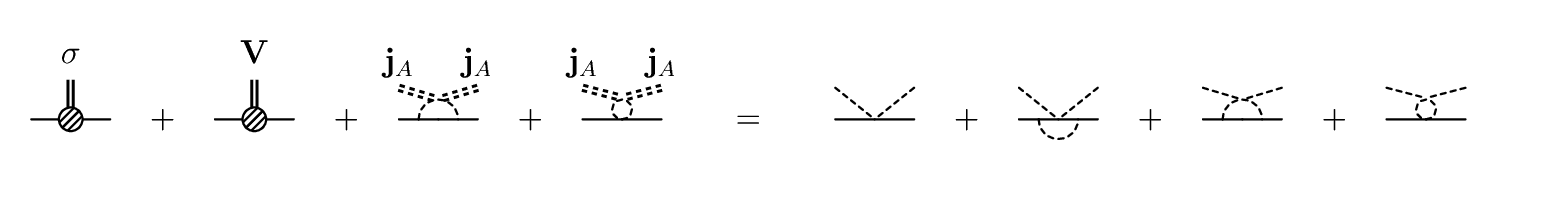}
\end{center}
\end{figure}

\noindent 
with the left-hand side containing the proper coefficient given by the
Ward identity. The other possible graphs from the pion calculation are in
one-to-one correspondence with the graphs of the same topology from
the two axial-vector correlator.

\end{document}